\documentclass[12pt]{article}

\usepackage{bm}
\usepackage{float}

\usepackage{graphicx}
\usepackage{amsmath}
\usepackage{upgreek}
\usepackage[sort&compress]{natbib}

\newcommand{\bea}{\begin{eqnarray}}
\newcommand{\eea}{\end{eqnarray}}

\usepackage{times}
\usepackage{mathrsfs}
\usepackage{graphicx}

\usepackage{caption}

\begin{document}

\title{\vspace{-1.5cm} \large \textbf{Fano effect in an ultracold atom-molecule coupled system} \vspace{-0.5cm}}
\author{
\small Yuqing Li$^{1}$, Guosheng Feng$^{1}$, Jizhou Wu$^{1,4}$, Jie Ma$^{1,4,*}$, Bimalendu Deb$^{2,3,*}$, Arpita Pal$^{2}$, \vspace{-0.3cm}\\
\small Liantuan Xiao$^{1,4}$ and Suotang Jia$^{1,4}$ \vspace{-0.3cm}\\
\small $^1$State Key Laboratory of Quantum Optics and Quantum Optics Devices, Institute of Laser spectroscopy, \vspace{-0.3cm}\\
\small College of Physics and Electronics, Shanxi University, Taiyuan 030006, China \vspace{-0.3cm} \\
\small $^2$Department of Materials Science, Indian Association for the Cultivation of Science, Jadavpur, \vspace{-0.3cm}\\
\small Kolkata 700032, India \vspace{-0.3cm}\\
\small $^3$Raman Centre for Atomic, Molecular and Optical Sciences, Indian Association for the Cultivation \vspace{-0.3cm}\\
\small of Science (IACS), Jadavpur, Kolkata 700032, India \vspace{-0.3cm}\\
\small $^4$Collaborative Innovation Center of Extreme Optics, Shanxi University, Taiyuan, Shanxi 030006, China \vspace{-0.3cm}\\
\small *mj@sxu.edu.cn and msbd@iacs.res.in \vspace{-1.2cm}\\
}
\date{}
\maketitle



\begin{abstract}
\normalsize{The Fano effect or Fano resonance with a characteristically asymmetric line shape originates from quantum interference between direct and indirect transition pathways in continuum-bound coupled systems, and is a ubiquitous phenomenon in atomic \cite{Madden1963, Fano1968}, molecular \cite{Linn1983}, nuclear \cite{Feshbach1958, Feshbach1962} and solid-state physics \cite{Kroner2008, Fan2010, Schmidt2010}. In optical nanoscale structures,  the Fano effect has wide-ranging applications that include optical filtering \cite{Liu2009}, sensing \cite{Chen2009}, all-optical switching \cite{Yanik2003}, quantum interferometry \cite{Kobayashi2002} and nonlinear optics \cite{Samson2010}, and this opens new avenues for photonic devices. The emergent area of ultracold atomic and molecular gases presents an ideal platform for studying Fano resonances, since the physical parameters of these gases can be extensively tuned with high precision using external fields. However, an experimental demonstration of the Fano effect in hybridized atom-molecular coupled systems has  remained elusive. Here, we report on observations of the Fano effect in molecular spectra obtained by  photoassociation (PA) near a $\emph{d}$-wave Feshbach resonance. This effect occurs due to quantum interference in PA transitions involving the continuum of atom-atom scattering states, the underlying Feshbach and photoassociated excited bound molecular states. We measure the variation in atom loss rate with an external magnetic field close to the Feshbach resonance in the presence of PA laser, and thereby clearly demonstrate the Fano effect. Our results further reveal that the Fano effect has significant influence on spectral shifts. Based on Fano's method, we develop a theory that explains the observed experimental results relatively well. Our theoretical formulation takes into account quantum interference between or among multiple transition pathways and between inelastic channels. Our results present a novel method for tuning the collisional interaction strength with laser light using Fano resonance.}
\end{abstract}

\setlength{\parindent}{2em}

\section*{Introduction}
Resonantly coupled physical systems exhibiting dispersive and nontrivial line shapes have been attracting growing attention in recent years. In spectroscopy, the symmetric Lorentzian line shape is one of the most ubiquitous spectral features observed in fluorescence and absorption. It corresponds to an exponentially decaying excited state with a finite lifetime. In contrast, the asymmetric Fano line shape emerges when quantum interference takes place between two competing optical transition pathways, one connecting a bound state embedded in a continuum of states with a discrete state, and the other connecting the discreet state  with the continuum. In order to explain the experimentally observed asymmetric line shapes in the absorption of noble gases \cite{Beutler1935}, Fano presented an important formula in 1961 for  scattering cross section, $\sigma=(\epsilon+\emph{q})^{2}/(\epsilon^{2}+1)$, involving the shape parameter $\emph{q}$ and the reduced energy $\epsilon$ defined by $2(\emph{E}-\emph{E}_{\emph{F}})/\Gamma$, where $\emph{E}_{\emph{F}}$ is the resonant energy and $\Gamma$ is the width of an autoionizing state \cite{Fano1961}. Although Fano resonance has been observed in many physical systems, including atomic \cite{Madden1963, Fano1968}, nuclear \cite{Feshbach1958, Feshbach1962} and solid-state physics \cite{Kroner2008, Fan2010, Schmidt2010} and molecular spectroscopy \cite{Linn1983}, this formula is still in use in a canonical form to describe the universal behavior of Fano resonance. The $\emph{q}$ parameter governs the symmetry of the line shape, and four distinct regimes can be identified: $\emph{q}$=$\infty$ corresponds to a Lorentzian line shape; $\emph{q}$=0 corresponds to an inverted symmetric Lorentzian line shape; and $\emph{q}>0$ and $\emph{q}<0$  correspond to asymmetric line shapes with anomalous and normal dispersion-like anti-symmetries, respectively. Systems supporting such resonant coupling are of great importance in terms of the potential applications of their characteristically asymmetric line shapes and the drastic variation of the asymmetry depending on the coupling conditions. In particular, the electromagnetically induced transparency (EIT) resulting from a Fano-like quantum interference is a key topic of research, with a plethora of applications in atomic physics and quantum optics, such as the slowing of light, quantum memory, enhanced frequency conversion, lasers without inversion, etc \cite{Harris1997, Fleischhauer2005}.

Over the years, Fano resonance has been studied in a wide variety of systems, such as semiconductors \cite{Lee2006}, quantum dots \cite{Kobayashi2002}, metamaterials \cite{Lukyanchuk2010}, photonic crystals \cite{Fan2002} and waveguide arrays \cite{Weimann2013}. In these optical nanoscale structures,  the Fano effect has found widespread  applications \cite{Liu2009, Chen2009, Yanik2003, Kobayashi2002, Samson2010} and this has opened up new research areas for photonic devices. A novel type of nonlinear Fano resonance has been found in coupled plasmonic-molecular systems \cite{Neubrech2008, Dregely2013}, in which a broadband plasmon interferes with the narrowband vibration of molecules. Such hybridized systems are useful in the precise control of line shapes, since the inherent properties of the system can be changed for each resonance and its optical response characteristics; this may be implemented not only through the choice of the structure of the matter, but also by introducing an external drive field that affects the matter. More recently, the all-optical control over Fano line shapes observed in a resonantly coupled plasmonic-atomic system \cite{Stern2014} has been shown to be important for developing all-optical switching. The coupling between plasmonic and atomic resonances produces spectral line shapes that cannot be obtained by any of the individual materials. It would therefore be of great interest to control the interplay between two resonances by simply changing the phase-matching conditions of the hybrid system. This degree of freedom allows the observation of the gradual evolution of a quantum coherent state \cite{Glauber1963}. However, most experiments on Fano resonance are usually performed in systems with uncontrollable autoionization or an analogous process through which an  excited state is coupled to a continuum. Furthermore, it is difficult to create a hybridized system with a capacity for the individual control of each resonance. In this context, an ultracold atom-molecule coupled system is very appealing for studying the Fano effect, since the underlying atomic and molecular states are amenable to coherent control. The Fano effect in the photoassociation (PA) of cold atoms in the presence of a magnetic Feshbach resonance has been theoretically predicted previously \cite{Deb2009jpb2, Deb2010jpb}. This effect has been shown to be useful for the creation of a bound state in a continuum \cite{Deb2014pra} and may thereby find important applications in the coherent manipulation of atom-molecule coupled systems. Although a Fano-like asymmetric spectral profile has been observed in PA \cite{Junker2008}, a magneto-optically tunable Fano effect in an atom-molecule coupled system is yet to be experimentally confirmed.

Here, we report observations of the Fano effect in an ultracold atom-molecule coupled system composed of ultracold Cs atoms in the continuum of scattering states and $\emph{d}$-wave Feshbach molecules, both of which can be resonantly coupled to an excited molecular state by the same PA light. The observation of the dependence of PA rate on the magnetic field around the Feshbach resonance enables us to obtain the Fano effect with characteristically asymmetric line shapes. In contrast to current theories about enhanced PA using Feshbach resonance \cite{Junker2008, Mackie2008}, the nonlinear Fano model that we develop here using coupled dual Fano-type quantum interferences can satisfactorily explain our experimental observations. This is accompanied by the substantial enhancement or suppression of PA spectral shift, depending on the magnetic field near the Feshbach resonance. The narrow width of $d$-wave Feshbach resonance induces a sensitive variation of the PA rate with magnetic field close to the Feshbach resonance. This understanding can help underpin the applications of emerging hybridized ultracold atom-molecule systems based on the observed Fano effects, such as
magneto-optical quantum interference \cite{Deb2010jpb}, nonlinear optics \cite{Samson2010} and manipulation for interactions of ultracold atoms.

\section*{Scheme for generating Fano resonance}
\noindent  Ultracold $^{133}$Cs atoms in the hyperfine state $\emph{F}=3, \emph{m}_{F}=3$ are prepared using three dimensional degenerated Raman sideband cooling (3D DRSC) \cite{Li2015LPL} and then loaded into a crossed optical dipole trap with trap frequencies $(\omega_{x},\omega_{y},\omega_{z})/2\pi$ = (63, 56, 83) Hz using the magnetic levitation technique (Fig. 1a). The uniform magnetic field $\emph{B}$ oriented along the $z$-axis firstly held at a value of 75 G, that is, relatively far from the $\emph{d}$-wave Feshbach resonance at 47.97 G. A thermal equilibrium is obtained  for 500 ms during which large three-body interactions lead to a strong loss for the trapped Cs atoms \cite{Weber2003, Li2015PRA}. At this point, the number and temperature of atoms are found to be $2.5\times10^{5}$ and 3.5 $\mu$K, respectively. The PA laser beam has a waist ($1/e^{2}$ radius of intensity) of 150 $\mu$m and is $\pi$ polarized in the same direction as $\emph{B}$. Prior to illuminating the PA laser on the atoms, $\emph{B}$ is adiabatically changed to a certain value near the $d$-wave Feshbach resonance \cite{Chin2000, Leo2000, Chin2004}. After illuminating for 100 ms, the laser and magnetic field are switched off simultaneously. By recording a series of PA spectra at different values of $\emph{B}$ around the Feshbach resonance location, we can investigate the Fano effect in a hybridized ultracold atom-molecule system. In order to distinguish the influence of the three-body loss in Feshbach resonance on atom loss from the two-body inelastic loss in PA, we also measure the atom loss induced by the Feshbach resonance without PA, and find that it can be ignored in our experiment. This can be attributed to the narrowness of the Feshbach resonance \cite{Chin2000, Chin2004} and the diluteness of the atomic gas in the thermal equilibrium process.

The basic level scheme used in our experiment is shown in Fig. 1b and c. The PA laser is tuned  near to PA resonance with a continuum-bound transition, where ``continuum'' refers to the scattering states of two electronically ground-state atoms, and the bound states belong to the outer well of an excited molecular $0_{g}^{-}$ potential below the $6S_{1/2}+6P_{3/2}$ threshold for Cs$_{2}$ (Fig. 1b). Owing to the small ($\sim 1 $ MHz) hyperfine splitting of a rovibrational level \cite{Danzl2010}, multiple bound states can be optically coupled by the same PA laser (Fig. 1c). In our model, we consider two such bound states  $\mid b_1 \rangle$ and $\mid b_2 \rangle$, both of which belong to the outer well of the excited molecular $0_{g}^{-}$ potential. Since $\emph{B}$ is tuned to around the energy level crossing point between the continuum $\mid E \rangle $ in the open channel $| g \rangle$ and a $d$-wave bound state $\mid b_{c} \rangle$ in the closed channel, $\mid c \rangle$ becomes embedded in the continuum $| E \rangle$. Here, $E$ refers to the collision energy. Due to the small energy difference between  $| E \rangle$ and $|b_{c}\rangle$, the PA laser with a frequency near-resonant with the continuum-bound transitions $| E \rangle \leftrightarrow |b_1 \rangle$ and $| E \rangle \leftrightarrow |b_2 \rangle$ simultaneously induces the bound-bound transitions $|b_{c}\rangle \leftrightarrow |b_1 \rangle$ and  $|b_{c}\rangle \leftrightarrow |b_2 \rangle$.  Due to the short lifetime, the populations in $| b_1 \rangle$  and $| b_2 \rangle$ undergo spontaneous radiative decay predominantly into pairs of hot atoms that escape from the trap. By recording the atom loss for different values of $\emph{B}$ near the $d$-wave Feshbach resonance, we can obtain a series of PA spectra. Since both $\mid b_1 \rangle$ and $\mid b_2 \rangle$ are driven by the optical transitions under the same PA laser, there is a laser-induced
coupling between $\mid b_1 \rangle$ and $\mid b_2 \rangle$. As a result, both of these excited states can decay spontaneously to a state, giving rise to vacuum-induced or spontaneously generated coherence. We include these coherence effects in our theoretical formulation.

Figure 1d shows a two-dimensional spectrum of atom loss strength as a function of PA laser frequency $\omega_{L}$, which is tuned near the resonance to the $v$ = 10 vibrational level in the presence of a variable $B$ around the $\emph{d}$-wave Feshbach resonance. For the fixed detuning of a PA laser close to the resonant PA transition with the $J$ = 0 and $J$ = 2 rotational levels, the appearance of maximum loss in the number of atoms at some certain $B$ close to the Feshbach resonance location is consistent with the Feshbach-optimized PA \cite{Junker2008}. The spectrum for $J$ = 0 clearly shows an asymmetric variation in atom loss strength with $B$ around the point where the major peak (red) represents the maximum loss for atoms in the trap. In comparison, the fast variation of atom loss with $B$ leads to a subtle asymmetry for $J$ = 2. At a value of $B$ close to the point of maximum atom loss, there are some asymmetric changes of atom loss with PA laser detuning for both $J$ = 0 and $J$ = 2. These asymmetry characteristics,
 which can not be explained by the theory of enhancing PA using Feshbach resonance \cite{Pellegrini2008}, indicate the Fano effect.

\begin{figure*}
\centering
\includegraphics[width=14.cm, angle=0]{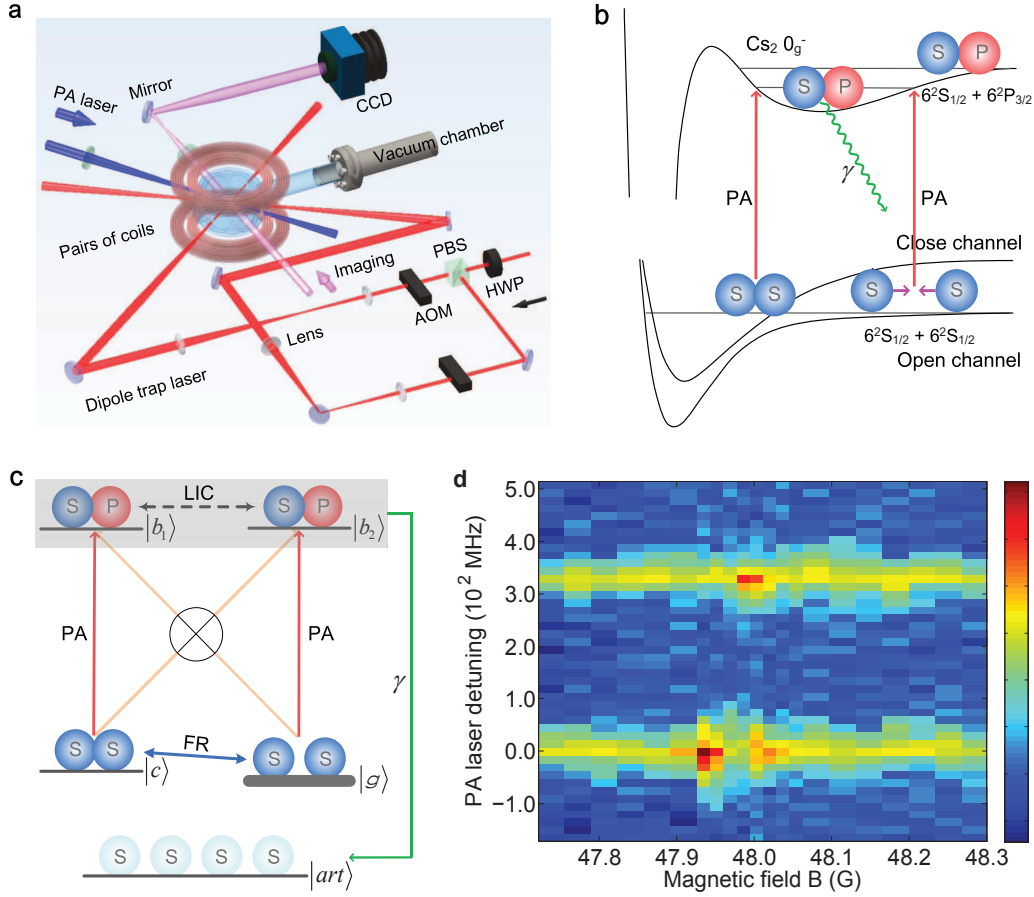}
\caption{\textbf{Experimental apparatus, schematic diagram of levels and atom loss spectrum.} \textbf{(a)} Experimental geometry and laser configuration. The PA laser is focused on the atoms trapped in the crossed dipole trap consisting of two horizontally crossing 1064 nm laser beams at an angle of 90$^{0}$. Pairs of quadrupole coils and Feshbach coils are used to produce the magnetic field gradient $\partial B$/$\partial z$ and the uniform bias field $ \emph{B}$ for the realization of magnetically levitated loading of Cs atoms and $d$-wave Feshbach resonance, respectively. \textbf{(b)} A schematic diagram of molecular potential. A pair of Cs atoms in the open channel is coupled to a $\emph{d}$-wave bound molecular state in the closed channel $| c \rangle$ via Feshbach resonance. The PA laser near-resonance with the free-bound transition can also induce the bound-bound transition, and forms electronically excited molecular bound states in the outer well of the $0_{g}^{-}$ potential below the $6S_{1/2}+6P_{3/2}$ threshold for Cs$_{2}$.  \textbf{(c)} Scheme for coupled dual Fano resonances. The same PA laser can induce couplings of both $\mid E \rangle$ and $\mid b_{c}\rangle$ to two quasi-degenerate excited molecular states $\mid b_1 \rangle$ and $\mid b_2 \rangle$ having the same vibrational and rotational quantum numbers but different molecular hyperfine quantum numbers. Thus, there are coupled two Fano resonances: one is from the quantum interference between PA transitions $\mid E \rangle$ $\leftrightarrow$ $\mid b_1\rangle$ and $\mid b_{c} \rangle$ $\leftrightarrow$ $\mid b_1\rangle$, and the other is from the quantum interference between PA transitions $\mid E \rangle$ $\leftrightarrow$ $\mid b_2)\rangle$ and $\mid b_{c} \rangle$ $\leftrightarrow$ $\mid b_2\rangle$. This leads to a laser-induced coupling (LIC) between $\mid b_1\rangle$ and $\mid b_2\rangle$. On the other hand, both $\mid b_1\rangle$ and $\mid b_2\rangle$ can spontaneously decay to a decay channel, representing a PA loss. \textbf{(d)} Two-dimensional spectra of atom loss strength. The number of atoms remaining in the trap after PA is recorded as a function of PA laser frequency at different $\emph{B}$ around the $d$-wave Feshbach resonance. The colormap is inverted to show that the PA-induced trap loss increases with a shift from blue to red.}
\label{scheme}
\end{figure*}

\section*{Observations of Fano effect in PA rate}

\noindent The loss rate $K_{E}$ at the collision energy $E$ and laser frequency $\omega_{L}$ on near-resonance is a function of $B$ near the $d$-wave Feshbach resonance (Figs. 2 and 3). As a dimensionless energy parameter, $\varepsilon$ is defined as the reduced energy in the Fano theory, and depends on the binding energy $E_{c}$ of the $d$-wave bound molecular state $| b_{c} \rangle$ and the width $\Gamma_{f}$ of Feshbach resonance. The Fano effect is most clearly manifested in the variation of $K_{E}$ with $B$. Thermally averaged loss rate $K_{av}$ is deduced from the time evolution of the atom number density $n(t)$, given by $\dot{n}(t)$=-$K_{av}$ $n^{2}(t)$. The PA process mainly determines the local atom loss compared to the less important three-body loss, as demonstrated in our experiment, and any possible time dependence of $K_{av}$ during the PA process is averaged over. For both excited rovibrational bound states with ($v=10$, $J=0$) and ($v=17, J=0$), our experimental results clearly exhibit Fano-type spectral features as shown in Fig. 2. However, unlike the standard Fano profile with one maximum and one minimum (the minimum is known as the Fano minimum), these spectra have prominent maxima with an asymmetric shape and an additional smaller peak. In other words, there are two Fano resonances: one has a strongly asymmetry curve with Fano parameter $| q_{1} |$ $\leq$ 1 and the other with $| q_{2} |$ $\gg$ 1. Using the coupled dual Fano resonances along with the hyperfine structure of the excited molecular level, we have developed an analytical theory to fit the experimental data for $J = 0$ with the relatively weak bound-bound transition limited by selection rules. According to the expression for the probability amplitude of excited bound state $| b_{n} \rangle$, the Fano minimum appears on the right side of Fano maximum when $q_{n}$ = -$\varepsilon$. We thus determine the two Fano parameters $q_{1}$ and $q_{2}$ by combining the theoretical analysis of the observed asymmetric spectrum in each of Figs. 2a and b. The parameters $\Gamma_{1}$ and $\Gamma_{2}$ reflect the free-bound transition strengths $| E \rangle$ $\rightarrow$ $| b_{1} \rangle$ and $| E \rangle$ $\rightarrow$ $| b_{2} \rangle$, respectively. These quantities are related to excited molecular rovibrational levels and the PA laser intensity in the experiment. The large difference between $\Gamma_{1}$ and $\Gamma_{2}$ indicates that there is a pronounced transition from the scattering state $| E \rangle$ to the bound state $| b_{1} \rangle$ with a weak but significant coupling between $| E \rangle$ and $| b_{2} \rangle$, allowing the observation of both the maximum and the second peak in $K_{av}$, as shown in Figs. 2a and b.

\begin{figure*}
\centering
\includegraphics[width=0.7\linewidth, angle=0]{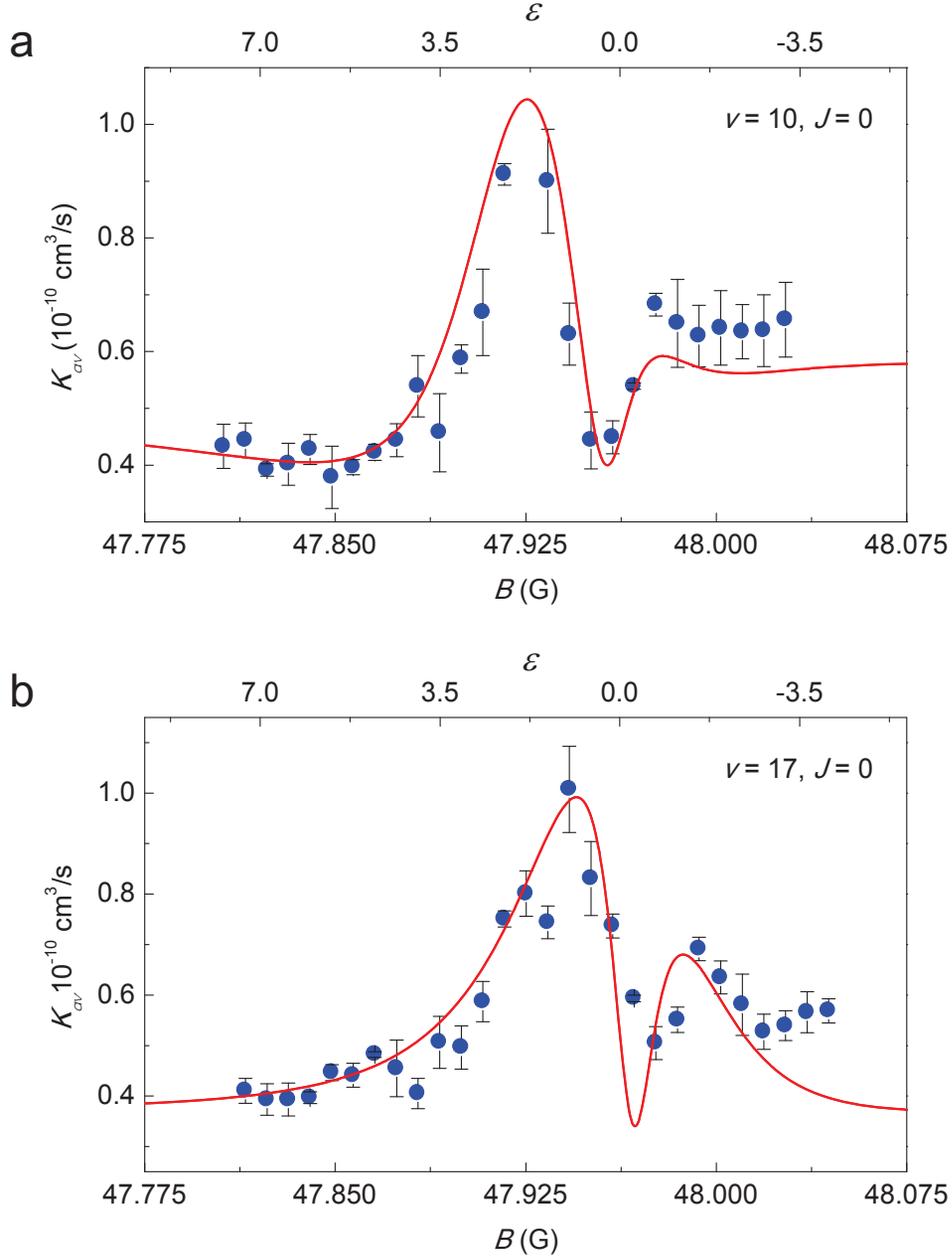}
\caption{The $K_{av}$ (spheres with error bars) as a function of $B$ near the $d$-wave Feshbach resonance for the rovibrational levels $v=10, J=0$ in \textbf{a} and for $v=17, J=0$ in \textbf{b}. By combing the observed Fano minimum in $B$ with other chosen parameters (see the text), we set $q_{1}$ = -0.3 and $q_{2}$ = 21.69 (\textbf{a}) and $q_{1}$ = 0.3 and $q_{2}$ = 21.69 (\textbf{b}). The solid line shows the prediction of the analytical theory from the coupled dual Fano resonances model described in the Methods section with $\Gamma_{1}$ = 15.5 MHz and $\Gamma_{2}$ = 0.04 MHz (\textbf{a}) and $\Gamma_{1}$ = 10.5 MHz and $\Gamma_{2}$ = 0.001 MHz (\textbf{b}).}
\label{paratev10}
\end{figure*}

To explain the observed experimental results, we develop a model using two ground-state channels and two excited channels, and obtain an analytical expression for the PA rate (see the Methods section). We fit the experimental data with the analytical formula using realistic and judiciously chosen values of the parameters of our model following the early works on PA \cite{Kraft2005, Lignier2011, Tolra2003} and Feshbach resonance \cite{Chin2000, Leo2000, Chin2004, Mark2007, Lange2009, Chin2010} of Cs atoms. Multi-channel calculations of magnetic Feshbach resonance, using up-to-date and high-precision molecular potential data from Cs$_2$, require at least 24 hyperfine channels to obtain agreement with the observed magnetic Feshbach resonances \cite{Chin2004}. Therefore our model represents a highly simplified situation of the actual physical scenario. The $d$-wave Feshbach resonance used in our experiment occurs close to the threshold of the lowest hyperfine channel \cite{Leo2000, Chin2004}. The Feshbach coupling $V_E$, which is related to the Feshbach resonance line width $\Gamma_f = 2 \pi |V_E|^2$, is a function of the collision energy $E$. Since the Feshbach resonance occurs at a very low energy ($< 5 \mu$K), which is almost close to the threshold, we can expect that $\Gamma_f(E)$ as a function of $E$ has a maximum at an energy $E \ne 0$, since  in the limit $E \rightarrow 0$ we have $V_E \rightarrow 0$. Like $\Gamma_f$, the other two free-bound PA stimulated line widths $\Gamma_1$ and $\Gamma_2$ are also energy-dependent. In the fitting of our model to the experimental data, we have assumed that $\Gamma_f$, $\Gamma_1$ and $\Gamma_2$ are weakly energy-dependent near the energy where the Feshbach coupling is a maximum. We have therefore taken these parameters as energy-independent fitting parameters. Another important parameter of our model is the spontaneous linewidth, which we fix at $\gamma_1 = \gamma_2 = \gamma_{sp} \simeq 17 $ MHz \cite{Kraft2005}.

In our theoretical calculation, if we ignore the hyperfine effect on the excited molecular rovibrational state and the cross-coupling between the two hyperfine states, we can then recover the standard Fano profile with a prominent maximum and minimum governed by only one parameter $q_f$ which, in the present context, can be called the Feshbach-Fano asymmetry parameter \cite{Deb2009jpb2}. For a standard Fano resonance, when $q$ is close to $\pm$1, both competing transitional pathways are of similar strength, and the line shape becomes very asymmetric. Especially for the excited molecular rovibrational level $v=10, J=0$, we have found that the PA laser  intensity should be comparable to the saturation intensity determined by the $\gamma_{sp}$ in order to get good fit. At low energy, the continuum-mediated effective coupling $V_{eff}$ between $\mid b_{c} \rangle$ and $\mid b_{1} \rangle$ or $\mid b_{2} \rangle$ is negative (see the Methods section). The expression for $q_f$ shows that it is positive if $\hbar \Omega > |V_{eff}|$, where $\Omega$ is the Rabi frequency for the bound-bound optical coupling between $\mid b_{c} \rangle$ and $\mid b_{n} \rangle$ ($n = 1, 2$). For $v=10, J=0$ (Fig. 2a), $q_{1}$ $<$ 0 corresponds to a relatively weak direct coupling $\hbar \Omega < |V_{eff}|$ compared to that for $v=17, J=0$ with $q_{1}$ $>$ 0 (Fig. 2b). In Fig. 2a and b,  a large value of $q_{2}$ ($ >\!> 1$) with small value of $\Gamma_{2}$ leads to a relatively small difference between the second peak and the background spectrum, where  $\Gamma_{2}$  is about two orders of magnitude lower than $\Gamma_{1}$. Despite $\Gamma_{2}$ being small, the experimental results with $q_{2}$ $>\!>$ 0 demonstrate that there is an effective bound-bound coupling. Previously, the free-bound Franck-Condon (FC) principle has been used to qualitatively explain the enhancement of the PA rate near $s$-wave Feshbach resonance \cite{Junker2008}; however, it is difficult to give  correct  predictions for the asymmetric shape and Fano minimum using the FC principle only.

To further understand how the Fano parameters $q_1$ and $q_2$ reflect the share of each of two competing transitions $| b_{c} \rangle \rightarrow | b_{n} \rangle$ and $| E \rangle \rightarrow | b_{n} \rangle$, we measured $K_{av}$ as a function of $B$ for the molecular rovibrational level $v$ = 10, $J$ = 2, as shown in Fig. 3, with  $J$ being different from that in Fig. 2a. Due to the $d$-wave bound molecular state $| b_{c} \rangle$ with the rotational quantum number $\ell$ = 2 in the closed channel $| c \rangle$, the selection rules allow for relatively strong bound-bound coupling  between the states $| b_{c} \rangle$ and $| b_{n} \rangle$ for $J$ = 2.  In comparison, the experimental data in Fig. 3 give a larger $q_{1}$ = 3.37 than that in Fig. 2a, with the large $\Omega$ resulting from the substantial bound-bound transitions between the states $| b_{c} \rangle$ and $| b_{1} \rangle$ for $J$ = 2. As is well-known from Fano theory, the parameter $q_{1}$ is consistent with the asymmetry of the observed spectrum that is more prominent for $J=0$ than that for $J=2$. An increased $\Omega$ leads to a weaker free-bound transition, which is reflected by the value of $\Gamma_{1}$ compared to that in Fig. 2a. The combination of the small peak resulting from a smaller $\Gamma_{2}$ with the Fano minimum corresponding to $q_{1}$ $>$ 1 forms the right side of asymptotic curve concentrating on a constant. The linewidth determined by the profile of the loss rate in Fig. 3 is very similar to the linewidth $\Gamma_{f}$ of the $d$-wave Feshbach resonance. This narrow linewidth also indicates the dominating effect of bound-bound transitions in the Fano resonances observed in our system.

\begin{figure*}
\centering
\includegraphics[width=0.7\linewidth, angle=0]{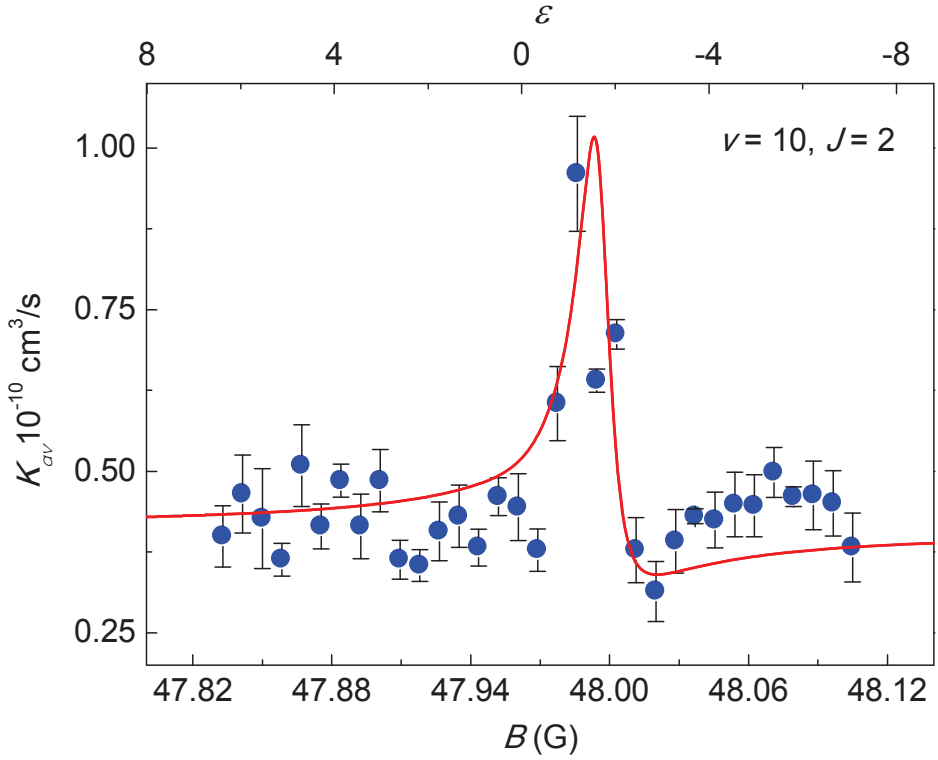}
\caption{Same as in Fig. 2a but for the rovibrational level $v=10, J=2$. The Fano parameters $q_{1}$ = 3.37 and $q_{2}$ = 7.82 are determined from the observed Fano minimum. The parameter $q_{1}$ $>$ 1 implies the predominant bound-bound transition $| b_{c} \rangle$ $\rightarrow$ $| b_{1} \rangle$ for $J$ = 2 compared to that for $J$ = 0 in Fig. 2a. The solid line is obtained using the coupled dual Fano resonance model with the experimental parameters $\Gamma_{1}$ = 6.2 MHz and $\Gamma_{2}$ = 0.3 MHz.}
\label{paratev17}
\end{figure*}

\section*{PA spectral shift}
\noindent The Fano effect is also evident in the PA laser intensity-induced frequency shift (also PA spectral shift). We have also measured $K_{av}$ in PA spectrum as a function of PA laser detuning $\delta$ for a series of $B$ near the $d$-wave Feshbach resonance. As $B$ varies around the Feshbach resonance, the position at which the maximum $K_{av}$ occurs is shifted. The location of a PA resonance at a particular $B$ is determined by fitting the resonance line shape in the PA spectrum to a Lorentzian. By recording PA spectra at several values of PA laser intensity $I_{PA}$ and $B$ for the molecular rovibrational level $v$ = 17, $J$ =0, we have observed a linear variation in the PA spectral shift $E'_{shift}$ with $I_{PA}$ at a particular $B$. The slope of PA spectral shift is extracted and shown in Fig. 4 as a function of $B$. Far from the Feshbach resonance, the shift is red and is accompanied with negative $E'_{shift}$ = -0.75 MHz/(W/cm$^{2}$), which is in agreement with previous investigation \cite{Li2015ol}. As the FR is approached from both low and high fields, the slope shows a dispersive character. The $B$ at which the singular point of $E'_{shift}$ appears is close to the position of the Fano minimum in $K_{av}$ in Fig. 2b. Based on the theoretical PA spectral shift derived by our theory, we use the parameters in Fig. 2b to give the theoretical curve as shown in Fig. 4.

\begin{figure*}
\centering
\includegraphics[width=0.7\linewidth, angle=0]{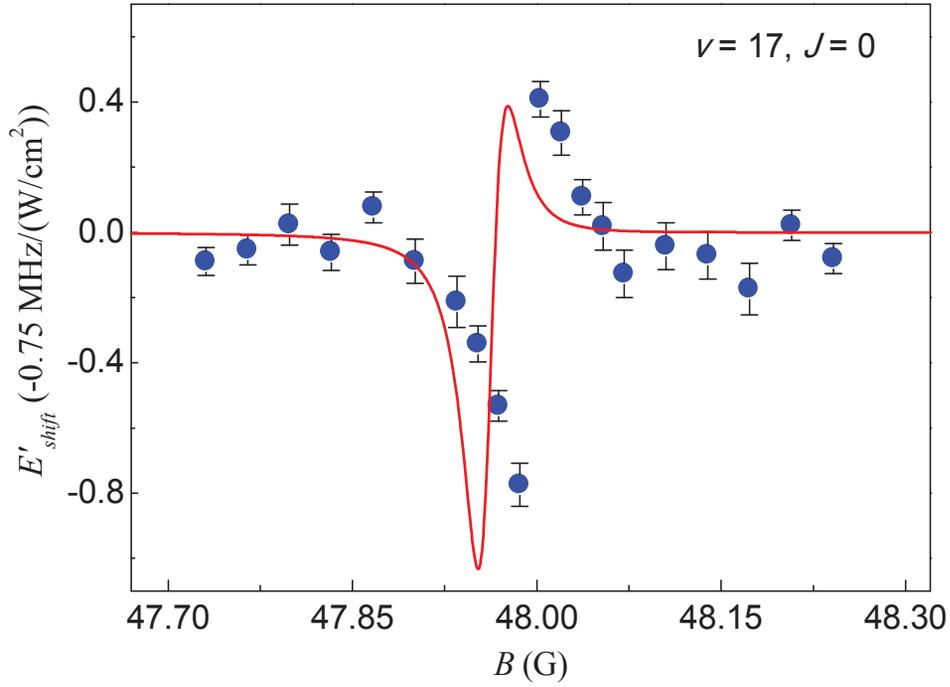}
\caption{The dispersive slope of PA spectral shift. The slope of PA spectral shift as a function of $B$ near the $d$-wave Feshbach resonance for the rovibrational level $v$ = 17, $J$ = 0. The error bar is the standard error of the fitted slope obtained by the laser intensity-dependent location of PA resonance at each $B$. The solid line is obtained by the theoretical calculation using the coupled dual Fano resonance model (see Methods section) with the parameters in Fig. 2b. The slope of PA spectral shift shows a dispersive manner as $B$ approaches the Fano minimum shown in Fig. 2b and go away it.}
\label{width}
\end{figure*}

\section*{Theory and  model}
\noindent In the presence of a magnetic Feshbach resonance, PA gives rise to  the Fano effect \cite{Deb2009jpb2}, which can dispersively modify the slope of PA spectral shift \cite{Deb2009jpb}. In our experiment,  we use a PA laser tuned near the rotational level $J=0$ or $J=2$ of the vibrational state  $v=10$ or $v=17$ in the excited molecular $0_g^{-}$ potential for Cs$_2$ below the dissociation (6S$_{1/2}$+6P$_{3/2}$). Here, the rotational angular momentum is defined by ${\mathbf J} = {\mathbf L} + {\mathbf S} + \vec{\ell}$, where ${\mathbf L}$ and ${\mathbf S}$ represent the molecular electronic orbital and spin angular momentum, and  $\vec{\ell}$  is the rotational angular momentum of relative motion of the two atoms that constitute the molecule. In our theoretical modeling, we consider hyperfine interaction in both the ground- and excited-state potentials. Then the  total angular momentum is $\vec{{\mathcal T}} = {\mathbf f} + \vec{\ell} = {\mathbf J} + {\mathbf I}$, where  ${\mathbf f} = {\mathbf F}_a + {\mathbf F}_b = {\mathbf J_e} + {\mathbf I}$, where  ${\mathbf F}_{a(b)}$ is the hyperfine angular momentum of the atom $a(b)$, ${\mathbf J}_e = {\mathbf L} + {\mathbf S}$ is the total electronic angular momentum and ${\mathbf I}$ is the total nuclear spin of the two atoms. In the presence of an external magnetic  field $B$, ${\mathcal T}$ is not a good quantum number; however, its projection $M = m_f + m_{\ell}$ on the $z$-axis  remains a good quantum number. However, at a low magnetic field, $f$ and $m_f$ can be considered approximately as conserved quantities.

Here we use a simplified  model of magnetic Feshbach resonance with two ground-state channels in the coupled molecular basis. One of the channels is open and the other is closed.  The $d$-wave Feshbach resonance used for our experimental demonstration of the Fano effect has been studied earlier, both experimentally and theoretically, by the groups of Chu and Grimm \cite{Chin2000, Leo2000, Chin2004, Mark2007, Lange2009, Chin2010}. In our two-channel modeling of this Feshbach resonance, the open channel $\mid g \rangle$ is asymptotically characterized by  $\mid (F_a=3, F_b=3) f=6, \ell=0; m_f=6 \rangle $ and the closed-channel $\mid c \rangle$ corresponds to $ \mid (F_a=3, F_b=3) f=4,\ell=2; m_f=4 \rangle $ in the absence of a magnetic field. The channel $\mid c \rangle$  supports a bound state $\mid b_c \rangle$. We assume that a single PA laser can couple several excited bound molecular states which are characterized by the quantum numbers $v, J, f, m_f$. For simplicity, we consider two excited bound states $\mid b_1 \rangle \equiv \mid (v, J) f_1, m_{f1} \rangle$ and $\mid b_2 \rangle \equiv \mid (v, J) f_2, m_{f2} \rangle$ being coupled to $\mid b_c \rangle$. Using the partial wave decomposition, the open channel bare scattering state $\mid E \rangle$ in the absence of any external field can be expressed as a superposition of partial-wave scattering states $\mid E, \ell m_{\ell} \rangle$. The state $\mid E, \ell m_{\ell} \rangle$ is coupled to $\mid b_c \rangle$  by the second order spin-orbit and spin-spin dipole interactions \cite{Chin2004}. To include spontaneous emission phenomenologically into our model, we assume that both $\mid b_1 \rangle$ and $\mid b_2 \rangle$ decay to the same artificial channel $\mid E''\rangle_{{\rm art}}$ as first introduced by Bohn and Julienne \cite{Bohn1999}. Let the interaction  of an excited bound state  with $\mid E''\rangle_{{\rm art}}$ be $ V_{{\rm art}}$. The relationship of the known spontaneous emission linewidth $\gamma_n$ with  $ V_{{\rm art}}$ is given by $\hbar \gamma_n = 2 \pi | _{{\rm art}} \langle E'' \mid V_{{\rm art}} \mid b_n \rangle |^2$. The loss of atoms due to this spontaneous emission at a particular collision energy $E$ is described by the loss rate $K_{{\rm E}}(\omega_{L}, B)$. Note that since both the excited bound states are optically driven by the same PA laser, there arises a laser-induced coupling between the bound states. On the other hand, since both of them can decay spontaneously to the same decay channel, which, in the present context, is modeled as an artificial channel following Bohn and Julienne \cite{Bohn1999}, there is a spontaneously generated coupling between them. The later one is also called vacuum-induced coupling or coherence \cite{Agarwal1974}, because the spontaneous emission is caused by vacuum fluctuations of the background reservoir of electromagnetic fields.

The model Hamiltonian describing PA in the presence of Feshbach resonance involving two ground-state channels,  two excited bound states and the artificial channel   is $H = H_0 + H_I $, where
\bea
H_0 &=& \int E d E \sum_{\ell,m_{\ell}} \mid E, \ell m_{\ell} \rangle \langle E, \ell m_{\ell} \mid
-  \sum_{n=1,2} \hbar \delta_n \mid b_n   \rangle \langle  b_n  \mid
\nonumber \\
&+& E_c \mid b_c \rangle \langle b_c \mid  +  \int E'' dE''  \mid E'' \rangle_{{\rm art}} \, _{{\rm art}} \langle E'' \mid
\eea
\bea
H_I &=& \sum_{\ell m_{\ell}}  \int d E  V_E^{\ell m_{\ell}} \mid b_c \rangle \langle E, \ell m_{\ell} \mid  + \sum_{n=1,2}  \Omega_n \mid b_c \rangle \langle b_n \mid   \nonumber \\
&+&  \sum_{n} \sum_{\ell m_{\ell}}  \int d E \Lambda_{n}^{ \ell m_{\ell}} (E) \mid b_n \rangle \langle E, \ell m_{\ell} \mid   \nonumber \\
&+&  \sum_{n}   \int d E'' V_{n,{\rm art}}(E'') \mid b_n \rangle \, _{{\rm art}}\langle E'' \mid + {\rm C.c.}
\eea
Here $\delta_n  = \omega_L - \omega_{b_n}$ is the detuning of the PA laser from the excited bound state frequency $\omega_{b_n}$, $E_{c}$ is the binding energy of the bound state $b_{c}$; $\Omega_n$ is the bound-bound Rabi frequency between $\mid b_c \rangle$ and $\mid b_n \rangle$, $V_E^{\ell, m_{\ell}} = \langle b_c \mid V \mid  E, \ell m_{\ell} \rangle$ is the coupling between $\mid E, \ell m_{\ell} \rangle$ and $\mid b_c \rangle$, where $V$ represents  the second order spin-orbit and spin-spin dipole interactions \cite{Chin2004, Mies1996}; $\Lambda_{n}^{\ell m_{\ell}}$ is the free-bound coupling between $|E,\ell m_{\ell} \rangle$ and $|b_{n} \rangle$, and $V_{n,{\rm art}}$ = $_{{\rm art}} \langle E'' |V_{{\rm art}}| b_{n} \rangle$. Following Fano's method, the Hamiltonian  can be diagonalized. Thus the dressed continuum state can be obtained:
\bea
\mid E,{\hat{k}} \rangle = &&\sum_{\ell' m_{\ell'}} Y_{\ell' m_{\ell'}}^{*}(\hat{k}) \left [ \sum_{n} A_{n}^{ \ell' m_{\ell'}}(E) \mid b_n \rangle   +  B_{E}^{\ell' m_{\ell'} }  \mid b_c \rangle  \right. \nonumber \\
&+& \left. \sum_{\ell m_{\ell}} \int d E' C_{E',\ell m_{\ell}}^{\ell' m_{\ell'}} \mid E',\ell m_{\ell} \rangle  + \int d E'' D_{E''}^{\ell' m_{\ell'}} \mid E'' \rangle_{{\rm art}}  \right ]
\label{eq3}
\eea
where $\hat{k}$ is an unit vector along the incident relative momentum of the two atoms, $B_E^{\ell' m_{\ell'}}$,  $A_{n}^{ \ell' m_{\ell'} }$, $C_{E',\ell m_{\ell}}^{\ell' m_{\ell'}}$  and $D_{E'' }^{\ell' m_{\ell'}}$ are the expansion coefficients which are explicitly derived in the Methods section.

It is the spontaneous decay from the dressed continuum  that leads to loss of the atoms in the trap. The $T$-matrix element for transition to the artificial decay channel is given by
\bea
T_{decay} =  \, _{\rm art}\langle E'' \mid V_{{\rm art}} \mid E, \hat{k} \rangle  = \sum_n  V_{n, {\rm art}} A_n^{\ell' m_{\ell'}}
\eea
The corresponding $S$-matrix element for the decay is given by
\bea
S_{decay} = - 2 \pi i T_{decay}
\eea
The total inelastic  scattering rate due to spontaneous emission is given by
\bea
\sigma_{inel} = \frac{ \pi }{k^2}  \mid S_{decay} \mid^2
\eea
The unitarity limit dictates that  $ \mid S_{decay} \mid^2  \le 1 $.
The loss rate at energy $E$ is given by
\bea
K_{E}(\omega_{L},B) =   | v_{rel} \sigma_{inel} |
\label{loss}
\eea
where $v_{rel}$ is the relative velocity related to the collision energy, which is given by $E = \mu v_{rel}^2/2 = \hbar^2 k^2/2\mu $. In fact, the observed loss rate has a narrow line, which means that only the low energy atoms ($E$ $<$ 3.5$k_{B}$ $\mu$K) contribute to the Fano effect.   We thus have (see the Methods section for details) the thermally averaged loss rate at the temperature $T$ near Feshbach resonance
\bea
K_{av}^{\rm res} ( {\rm \omega_{L}},B)
&=&  \frac{4 \pi^2 \hbar^2}{ \left ( 2 \pi \mu k_B T \right)^{3/2} }\int d E \exp\left (- \frac{E}{ k_B T } \right )  |S_{ decay}|^2
\label{lossE}
\eea
At low collision energy $E$, close to the threshold of the open channel,  $\left ( |S_{ decay}|^2 \right )$ is a slowly varying function of $E$.

Since the closed-channel bound-state energy is approximately a linear function of $B$ near the resonant value $B_{0}$, we expect that our formula (8) will give accurate results only near $B_{0}$. Furthermore, there may be off-resonant excitations of many other states not considered in our model can contribute to the loss. We therefore add a small off-resonant background contribution $K_{av}^{\rm bg}$ to obtain $K_{av}$ = $K_{av}^{\rm res}$ + $K_{av}^{\rm bg}$, which can be fitted for a considerable range of $B$ near $B_{0}$. We plot $K_{av}$ as function of $B$ for the fixed laser detuning $\delta_{n}$ and several other input parameters which are chosen based on the earlier works of the $d$-wave Feshbach resonance of Cs atoms \cite{Chin2000, Leo2000, Chin2004, Mark2007, Lange2009, Chin2010} in the hyperfine state $F = 3, m_{F} = 3$, and compare this with the experimental results. The Figs. 2 and 3 show good agreement between the experimental and theoretical plots. As discussed in detail in the Methods section, Eq. (\ref{lossE}) shows  that the loss spectrum will be given by the square of the coherent superposition of two Fano profiles $f_n = (q_n + \epsilon)/(\epsilon + i) $ with $n=1,2$, where $q_n$ is the well-known Fano asymmetry parameter and $\epsilon$ is the scaled energy detuning from the binding energy of the $d$-wave bound state. Each profile is modified by the cross coupling $Q_{12}$ between $\mid b_1 \rangle$ and $\mid b_2 \rangle$; this cross coupling arises due to simultaneous PA transitions to both  $\mid b_1 \rangle$ and $\mid b_2 \rangle$ from the ground-state manifold.  Note that the free-bound and bound-bound PA transitions predominantly occur at inter-atom separations of around 20 $a_{0}$, where $a_{0}$ is the Bohr radius. At such separations, several asymptotic channels can be mixed up by spin-orbit, spin-spin dipole and hyperfine interactions. In fact, earlier calculations \cite{Chin2004} show that one has to take into account at least 24 channels in order to get accurately quantitative results. Therefore, our model, which uses only two ground-state channels and two excited bound states, is an extremely simplified version of the actual physical situation. Nevertheless, we get reasonable agreement. Considering the incident ground-state channel (open) has  only $s$-wave ($\ell =0$) with $f=6$ and $m_f=6$, parity selection rules suggest that the excited bound states should have $f=7$ with $m_f=6$ (for $\pi$-polarized laser) or $m_f=7$ (for $\sigma_+$-polarized laser) or $f=5$ with $m_f=5$ (for $\sigma_{-}$-polarized laser). In our experiment, we apply a linearly polarized PA laser almost
perpendicularly to the magnetic field direction. Thus, the laser is predominantly of $\pi$-polarization; however, for a small misalignment of the laser propagation direction from the perpendicular to the magnetic field and a slight deviation in the polarization of PA laser, we have small component of circular polarizations as well. Therefore, if $\mid b_1 \rangle$ is chosen with $f=7$ and $m_f=6$, and $\mid b_2 \rangle$ with $f=5$ and $m_f=5$, the large peak of the spectrum will mostly correspond to contribution from $\mid b_1 \rangle$ and the small peak to $\mid b_2 \rangle$. We use the two asymmetry Fano parameters $q_1$ and $q_2$ as the fitting parameters determined by the location of the Fano minimum in the observed PA spectrum for the theoretical plots. In particular, for the result in Fig.3, since these two parameters are found to have positive values, this suggests that the bound-bound Rabi couplings $\Omega_1$ and $\Omega_2$ dominate over the corresponding free-bound couplings, in agreement with the predominant bound-bound transitions determined by the selection rules.

\section*{Summary and outlook}
\noindent
To demonstrate the Fano effect, we have measured the atom loss induced by PA near the $d$-wave Feshbach resonance and clearly observed not only enhanced but also suppressed PA associated with the maximum and minimum values in Fano-type asymmetric line shapes. Compared to the typical asymmetric line shape observed previously in Fano resonance in many physical systems, an additionally weak but obvious peak is observed here as a second maximum in addition to the usual Fano maximum.  In order to explain this experimental result, we have developed a model of a coupled dual Fano resonances in a hybridized ultracold atomic-molecular system. As shown in Fig. 1c, there are two Fano-type quantum interferences in a more complicated coupling mechanism that shows much rich physics than the common Fano effect. Our theory shows good agreement with the experimental observation of $K_{av}$ as a function of the $B$ near the $d$-wave Feshbach resonance. The primary profile characterized by both the Fano maximum and minimum $K_{av}$ in Fig. 2 and 3 can be thought of as the effect  arising from the Fano resonance with the participation of $\mid b_1 \rangle$. The Fano resonance from the coupling to $\mid b_2 \rangle $ also gives rise to a second maximum and the following Fano background. The difference in the asymmetry of the Fano line shapes originates from the discrepancy of Fano parameters $q_{f_1}$ and $q_{f_2}$ and the difference in the PA couplings.  We find that the two $q_{1}$-parameters are of the order of unity for the rovibrational levels $v$ = 10, $J$ = 0 in Fig. 2a and $v$ = 17, $J$ = 0 in Fig. 2b, implying the corresponding spectral asymmetry is significant. For $v$ = 10, $J$ = 2, both $q_{1}$ and $q_{2}$ are positive and have the same order, and this indicates that the bound-bound couplings dominate over the free-bound PA couplings. The overall spectral features result from the coupled two Fano profiles. In addition, the Fano effect has an important influence on the PA spectral shift, which shows a dispersive effect near the $d$-wave Feshbach resonance. The Fano theory also gives reasonable agreement with the experimentally obtained values for the slope of PA spectral shift.

Finally, our future research will be devoted to the realization of the optical manipulation of Fano resonances in ultracold atom-molecule coupled systems.  Using Aulter-Townes splitting of a bound Feshbach molecular state caused by an external laser field, we can construct two narrow $\emph{d}$-wave Feshbach resonances on either sides of the original resonance location, and would obtain double Fano asymmetric lineshapes with a tunable splitting. In this way, we can alter the positions of Fano maximum and minimum simultaneously by adjusting the intensity and detuning of the external laser field. As a result, it can be  used not only for switching of Fano resonances but also for studying non-linear optics in atom-molecule coupled systems. Conversely, the Fano effect dominated by a significant Fano parameter can be regarded as a probe for detecting whether there are some minute differences between the original Feshbach resonance and the two resonances formed by Aulter-Townes splitting. At present, there exists no another method for investigating this difference, although the splitting of Feshbach resonance is reported previously \cite{Bauer2009}. Furthermore, our results show that the use of a narrow $d$-wave Feshbach resonance to modify PA spectrum facilitates for the occurrence of relatively stronger bound-bound coupling which in turn will be useful for obtaining Rabi oscillations \cite{Yan2013, Taie2016} in time domain in atom-molecule coupled systems.

\section*{Methods}

\subsection*{Experiment}

\textbf{Preparation of ultracold Cs atoms in the crossed dipole trap}. Initially, $^{133}$Cs atoms are cooled to a temperature of about 200 $\mu$K in a magneto-optical trap (MOT). To prepare the ultracold atomic sample, the magnetic field gradient-compressed and optical molasses-cooled Cs atoms are finally cooled to a temperature of $\sim$ 1.7 $\mu$K and spin-polarized to the lowest hyperfine ground state $|F=3, m_{F}=3$ using three-dimensional degenerated Raman sideband cooling \cite{Li2015LPL}. A magnetic field gradient of $\partial B / \partial z$ = 31.3 G/cm and crossed dipole trap are employed to levitate and collect the cooled atoms, as depicted in Fig. 1a, in which the gravity of the Cs atom induces a large anti-trapping potential in the vertical ($z$) direction \cite{Li2015PRA}. The dipole trap is formed by intersecting two 1064 nm laser beams with the 1/$e^{2}$ beam radii of 230 $\mu$m and 240 $\mu$m in the $x$ and $y$ directions, respectively. During the magnetically levitated loading process of the crossed dipole trap, an uniform magnetic field of $\sim$ 75 G in the $z$ direction is switched on to cancel the resulting anti-trapping potential in the horizontal direction from the application of the vertical magnetic field gradient. The number of atoms trapped in the dipole trap and the atomic density are measured by the standard absorption image along the horizontal direction. At the uniform field of $\sim$ 75 G, the corresponding $s$-wave scattering length $a$ = 1230 $a_{0}$ leads to a strong three-body loss for optically trapped Cs atoms \cite{Weber2003}. After 500 ms of the thermal equilibrium process dominated by this strong three-body loss whose rate is proportional to the third power of atom density \cite{Weber2003, Bloom2013}, the dilute atomic sample presents few responses in the atom number to the $d$-wave Feshbach resonance. Thus, we can attribute the variation in the number of atoms remaining in the dipole trap to the PA-induced loss, and deduce the atom loss rate near the $d$-wave Feshbach resonance.

\subsection*{Theory}

We briefly discuss here the method of derivation of the dressed continuum of Eq. (3). A detailed derivation is given in the Supplementary Information. In particular, we obtain an analytical expression of the coefficient $A_{n}^{\ell' m_{\ell'}}$,  which  is the  probability amplitude for excitation of $\mid b_n\rangle$ ($n=1,2$) when the incident partial-wave of relative motion of the two atoms is $\ell'$ and its projection along the space-fixed $z$-axis is $m_{\ell'}$. From the time-independent Schr$\ddot{o}$dinger equation $\hat{H}\mid E,{\hat{k}} \rangle = E \mid E,{\hat{k}} \rangle$, we obtain a set of five coupled algebraic equations for $A_{1}^{\ell' m_{\ell'}}$, $A_{2}^{\ell' m_{\ell'}}$, $B_{E}^{\ell' m_{\ell'}}$, $D_{E'' }^{\ell' m_{\ell'}}$ and $C_{E'\ell m_{\ell}}^{\ell' m_{\ell'}}(E)$. The partial-wave symbols appearing in the superscript and subscript of $C_{E' \ell m_{\ell}}^{\ell' m_{\ell'}}$ refer to the incident and scattered  partial waves, respectively. $C_{E'\ell m_{\ell}}^{\ell' m_{\ell'}}$ is required to fulfil the scattering boundary conditions at large separation of the two atoms. We first eliminate $C_{E'\ell m_{\ell}}^{\ell' m_{\ell'}}$ and $D_{E'' }^{\ell' m_{\ell'}}$ to obtain three coupled equations for the amplitudes of the three bound states of our model.
We can then explicitly solve these three coupled equations. We introduce the dimensionless energy parameter $\epsilon = \left (E - E_c - E_c^{{\rm shift}} \right )/(\Gamma_f/2)$ where $E_c^{{\rm shift }}$ is the shift of $\mid b_c \rangle$ due to the inter-channel coupling, and express $A_{n}^{\ell'm_{\ell'}} (E)$  in the form
\bea
A_{n}^{\ell' m_{\ell'}} = {\mathscr D}_n^{-1} \left [R_{n}^{\ell' m_{\ell'}} + \xi_{n'}^{-1} Q_{n n'} R_{n'}^{\ell' m_{\ell'}}    \right ], \hspace{0.5cm} n \ne n'
\eea
where
\bea
R_{n}^{\ell' m_{\ell'}} = \Lambda_{n}^{\ell' m_{\ell'}}(E) + (q_n-i)\pi\sum_{\ell m_{\ell}} \Lambda_{n}^{\ell m_{\ell}} V_E^{\ell m_{\ell}}
 \frac{V_E^{\ell' m_{\ell'}}}{\frac{\Gamma_f}{2}(\epsilon+i)},
\eea
$\Lambda_{n}^{\ell m_{\ell}}(E)$ is the free-bound coupling between $\mid b_n \rangle$ and the partial-wave bare scattering state $\mid E, \ell m_{\ell} \rangle$, and $V_E^{\ell m_{\ell}}$ is the coupling between $\mid E, \ell m_{\ell} \rangle$ and $\mid b_c \rangle$ due to the second order spin-orbit and spin-spin dipole interactions. The Feshbach resonance linewidth is given by $\Gamma_f = 2 \pi \sum_{\ell m_{\ell}} |V_E^{\ell m_{\ell}}|^2$ and the stimulated linewidth of PA is $\Gamma_n = 2 \pi \sum_{\ell m_{\ell}} |\Lambda_{n}^{\ell m_{\ell}}|^2$.  The Fano-$q$ parameter $q_n$ is defined by
\bea
q_n = \frac{\Omega_n + V_{n,eff}}{\pi\sum_{\ell m_{\ell}} \Lambda^{\ell m_{\ell}}_{n}V_{E}^{\ell m_{\ell}}}
\eea
where
\bea
V_{n, eff}(E) = \sum_{\ell m_{\ell}} {\cal P}  \int dE' \frac{V^{\ell m_{\ell}}_{E'} \Lambda_{\ell m_{\ell}}^{n}(E')}{E-E'}
\eea
is the effective coupling between $\mid b_c \rangle$ and $\mid b_n \rangle$ mediated through the open channel continuum.

Here
\bea
{\mathscr D}_n = E + \hbar \delta_n - E_{n,pa}^{{\rm shift}} - E_{q n}^{{\rm shift}} + i \hbar ( \gamma_n + \Gamma_{q n})/2  - \xi_{n'}^{-1} Q_{n n'} Q_{n' n}.
\eea
with $\Gamma_{qn} = \Gamma_n - 2 {\mathrm Im} {\mathscr E}_{q n}$ and $ E_{q n}^{{\rm shift}} = {\mathrm Re} {\mathscr E}_{q n}$ where
\bea
{\mathscr E}_{q n} = \frac{\left((q_n-i)\pi\sum_{\ell m_{\ell}} \Lambda_{n}^{\ell m_{\ell}} V_E^{\ell m_{\ell}}\right)^2 }
{\frac{\Gamma_f}{2} (\epsilon + i)}.
\eea
The PA laser shift of $\mid b_n \rangle$ in the absence of magnetic field is
$E_{n,pa}^{\rm{shift}} = \sum_{\ell m_{\ell}} {\mathcal P}  \int dE' \frac{|\Lambda^{\ell m_{\ell}}_{n}(E')|^2} {E - E'}$.
The quantity
\bea
Q_{n n'} = \sum_{\ell m_{\ell}} \int dE' \frac{\Lambda^{\ell m_{\ell}}_{n}(E') \{\Lambda^{\ell m_{\ell}}_{n'}(E')\}^*}{E-E'}
+ \left((q_n-i)\pi\sum_{\ell m_{\ell}} \Lambda_{n}^{\ell m_{\ell}} V_E^{\ell m_{\ell}}\right)
\times  \frac{\pi\sum_{\ell m_{\ell}} \Lambda^{\ell m_{\ell}}_{n'}V_{E}^{\ell m_{\ell}}(q_{n'}-i)}{\frac{\Gamma_f}{2} (\epsilon + i)}
\eea
is the cross coupling between the two excited bound states $\mid b_1 \rangle$ and $\mid b_2 \rangle$.  The first term on the right-hand side of Eq.(13) arises due to the laser-induced coupling, and the second is caused by the vacuum-induced effect.
As a result of these couplings, the spontaneous emissions from the two excited bound states become correlated. Thus, it is the spontaneous decay from the dressed continuum or effectively from a coherent superposition state of the two bound states that leads to a loss of the atoms from the trap.
The $T$-matrix element $T_{decay}$ given by Eq. (4) in main text can thus be explicitly calculated and so is the resonant rate
$K_{av}^{{\rm res}}$ given in Eq.(8) in main text.
Since the $s$-wave contribution  is most significant near the threshold, we can use $V_E^{00} \simeq \sqrt{\Gamma_f/(2 \pi)} $ and $\Lambda_n^{0 0} \simeq \sqrt{\Gamma_n/(2 \pi)}  $. Then  $R_n^{0 0}$ reduces to the standard Fano profile form $F_n = \frac{\epsilon + q_n}{\epsilon+i}$. Clearly, $K_{av}^{{\rm res}}$ is given by a coherent superposition of two Fano profiles; each profile is modified by the cross coupling $Q_{12}$ between the two excited bound states. Thus, it can be expected that there will be two peaks and one or two minima. No peak exceeds the unitarity limit $S_{decay} \rightarrow 1$.
If we assume that the quantity $\left ( |S_{decay}|^2 \right )$ is a slowly varying function of $E$ then we can approximate the integral over the Maxwell-Boltzmann velocity distribution and write
$
K_{PA}^{{\rm res}} \simeq   k_B T/(h Q_T)  |S_{decay}|^2
$
where $Q_T = \left ( \frac{ 2 \pi \mu k_B T}{h^2} \right)^{3/2} $


\section*{Acknowledgements}

We would like to thank P. Zhang and D. Wang for helpful discussions. This research is supported by the State Key Development Program for Basic Research of China (Grant No. 2012CB921603), the Changjiang Scholars and Innovative Research Team in the University of the Ministry of Education of China (Grant No. IRT13076), National Natural Science Foundation of China (Grants No. 91436108, No. 61378014, No. 61308023, No. 61378015, No. 11402140, No. 11404197, and No. 11434007), Scientific Research Fund for the Doctoral Program of Higher Education of China (Grant No. 20131401120012).

\clearpage
\newpage
\begin{center}
\textbf{\Large Supplementary Information}\\*
\end{center}

\setcounter{equation}{0}
The model Hamiltonian describing photoassociation in the presence of Feshbach resonance involving one open channel, one closed channel,  two excited bound states and an artificial channel for the inclusion of spontaneous emission   is $H = H_0 + H_I $, where
\bea
H_0 &=& \int E d E \sum_{\ell,m_{\ell}} \mid E, \ell m_{\ell} \rangle \langle E, \ell m_{\ell} \mid
-  \sum_{n=1,2} \hbar \delta_n \mid b_n   \rangle \langle  b_n  \mid
\nonumber \\
&+& E_c \mid b_c \rangle \langle b_c \mid  + \int E'' dE''  \mid E'' \rangle_{{\rm art}} \, _{{\rm art}} \langle E'' \mid
\eea
\bea
H_I &=& \sum_{\ell m_{\ell}}  \int d E  V_E^{\ell m_{\ell}} \mid b_c \rangle \langle E, \ell m_{\ell} \mid  + \sum_{n=1,2}  \Omega_n \mid b_c \rangle \langle b_n \mid   \nonumber \\
&+&  \sum_{n} \sum_{\ell m_{\ell}}  \int d E \Lambda_{n}^{ \ell m_{\ell}} (E) \mid b_n \rangle \langle E, \ell m_{\ell} \mid   \nonumber \\
&+&  \sum_{n}   \int d E'' V_{n,{\rm art}}(E'') \mid b_n \rangle \, _{{\rm art}}\langle E'' \mid + {\rm C.c.}
\eea
where $\Omega_n$ is the bound-bound Rabi frequency between $\mid b_n \rangle$ and $\mid b_c \rangle$,  $V_E^{\ell, m_{\ell}} = \langle b_c \mid V \mid  E, \ell m_{\ell} \rangle$ is the coupling between the partial-wave ($\ell, m_{\ell}$) scattering state in the open channel and the bound state $\mid b_c \rangle$ in the closed channel. Here $\mid b_n \rangle$  represent the excited molecular bound state with vibrational quantum number $v$ and rotational quantum number $J_n$ with $M_{J_n}$ being the projection of $J_n$ on the space-fixed $z$-axis; $\mid E, \ell m_{\ell} \rangle$ is the energy-normalized scattering  state with partial wave $\ell$ and its projection $m_{\ell}$ on the space-fixed $z$-axis;  and $\delta_n  = \omega_L - \omega_{b_n}$ is the detuning of the laser from the excited bound state frequency $\omega_{b_n}$. Using Fano's theory, the Hamiltonian  can be exactly diagonalized. Thus one can obtain the dressed continuum state
\bea
\mid E,{\hat{k}} \rangle = &&\sum_{\ell' m_{\ell'}} Y_{\ell' m_{\ell'}}^{*}(\hat{k}) \left [ \sum_{n} A_{n}^{ \ell' m_{\ell'}}(E) \mid b_n \rangle   +  B_{E}^{\ell' m_{\ell'} }  \mid b_c \rangle  \right. \nonumber \\
&+& \left. \sum_{\ell m_{\ell}} \int d E' C_{E',\ell m_{\ell}}^{\ell' m_{\ell'}} \mid E',\ell m_{\ell} \rangle  + \int d E'' D_{E''}^{\ell' m_{\ell'}} \mid E'' \rangle_{{\rm art}}  \right ]
\label{eq3}
\eea
where $\hat{k}$ is a unit vector along the incident relative momentum of the two atoms, $B_E^{\ell' m_{\ell'}}$,  $A_{n}^{ \ell' m_{\ell'} }$, $C_{E',\ell m_{\ell}}^{\ell' m_{\ell'}}$
and $D_{E'' }^{\ell' m_{\ell'}}$
are the expansion coefficients. The superscript $\ell' m_{\ell'}$ indicates incident partial wave and the subscript $\ell m_{\ell}$ refers to the outgoing or scattered partial wave.  Physically, $A_{n}^{\ell' m_{\ell'}}$ implies probability amplitude for excitation of $\mid b_n\rangle$ when the incident partial-wave of relative motion of the two atoms is $\ell'$ and its projection along the space-fixed $z$-axis is $m_{\ell'}$. Similarly, $B_{E}^{\ell' m_{\ell'}}$ denotes the probability amplitude for the occupation of  $\mid b_c (\ell=2)\rangle$ for the $(\ell' m_{\ell'})$ incident partial wave. Similarly, $D_{E'' }^{\ell' m_{\ell'}}$ represents the probability amplitude of the artificial channel state $\mid E'' \rangle_{{\rm art}}$.

Here we present detailed derivation of the anisotropic dressed continuum  of Eq.(\ref{eq3}). From time-independent Schr$\rm \ddot{o}$dinger equation $\hat{H}\mid E,{\hat{k}} \rangle = E \mid E,{\hat{k}} \rangle$, we obtain the following set of coupled algebraic equations
\bea
(- \hbar \delta_n - E) A_{n}^{\ell'm_{\ell'}} + \Omega^{*}_n B_{E}^{\ell'm_{\ell'}} = &-& \sum_{\ell m_{\ell}} \int dE'\Lambda_{n}^{\ell m_{\ell}}(E') C_{E' \ell m_{\ell}}^{\ell'm_{\ell'}}(E)
- \int d E'' V_{n,{\rm art}}(E'') D_{E''}^{\ell' m_{\ell'}}(E)
\label{eq4}
\eea
\bea
(E_c - E) B_E^{\ell'm_{\ell'}} + \sum_{n=1,2} \Omega_n A_{n}^{\ell' m_{\ell'}} (E) = &-& \sum_{\ell m_{\ell}} \int dE' V_{E'}^{\ell m_{\ell}} C_{E' \ell m_{\ell}}^{\ell'm_{\ell'}}(E)
\label{eq5}
\eea
and
\bea
(E' - E) C_{E'\ell m_{\ell}}^{\ell' m_{\ell'}}(E) + \sum_{n=1,2} \{ \Lambda_{n}^{\ell' m_{\ell'}}\}^*(E') A_{n}^{\ell' m_{\ell'}} + V_{E'}^{\ell m_{\ell}} B_{E}^{\ell' m_{\ell'}} = 0
\label{eq6}
\eea
\bea
(E'' - E) D_{E''}^{\ell' m_{\ell'}} +  \{ V_{n, {\rm art}}^{\ell' m_{\ell'}}\}^*(E'') A_{n}^{\ell' m_{\ell'}}  = 0
\label{eq7}
\eea

Since the coefficient $C_{E'\ell m_{\ell}}^{\ell' m_{\ell'}}$ is required to fulfill scattering boundary conditions at large separation of the two atoms, we can express Eq. (\ref{eq6}) in the form
\bea
C_{E' \ell m_{\ell}}^{\ell' m_{\ell'}}(E) &=& \delta (E-E') \delta_{\ell,\ell'} \delta_{m_{\ell} m_{\ell'}} + \frac{V_{E'}^{\ell m_{\ell}}}{E-E'} B_{E}^{\ell' m_{\ell'}}\nonumber \\ &+& \sum_{n} \frac{ \{ \Lambda^{\ell m_{\ell}}_{n} \}^*(E')}{E - E'} A_{n}^{\ell' m_{\ell'}} (E)
\label{eqce}
\eea
The partial-wave symbols appearing in the superscript and subscript of   $C_{E' \ell m_{\ell}}^{\ell' m_{\ell'}}$ refer to the incident and scattered  partial waves, respectively. From Eq. (\ref{eq7}), we have
\bea
 D_{E''} (E) =  \sum_{n} \frac{\{ V_{n, {\rm art}}\}^*(E'') A_{n}^{\ell' m_{\ell'}} }{E - E''}
\label{eq77}
\eea

Substituting Eq. (\ref{eqce}) and Eq. (\ref{eq77}) in  Eqs. (\ref{eq4}) and (\ref{eq5}), we obtain
\bea
(- \hbar\delta_n - E) A_{n}^{\ell'm_{\ell'}} (E) + \Omega_n^* B_{E}^{\ell'm_{\ell'}} = &-& \Lambda_{n}^{\ell' m_{\ell'}}(E)  - \sum_{\ell m_{\ell}}\left [ \int dE' \frac{V^{\ell m_{\ell}}_{E'} \Lambda^{\ell m_{\ell}}_{n}(E')}{E-E'}\right] B_{E}^{\ell' m_{\ell'}} \nonumber\\
&-& \sum_{j=1,2} \sum_{\ell m_{\ell}}\left [ \int d E'
\Lambda^{\ell m_{\ell}}_{n}(E') \frac{ \{\Lambda^{\ell m_{\ell}}_{j}(E')\}^*} {E - E'} \right ]A_{j}^{\ell' m_{\ell'}}(E)
\nonumber \\
 &-&  \sum_{n'} \int d E'' \frac{   V_{n, {\rm art}}(E'') \{ V_{n', {\rm art}}\}^*(E'') }{E - E''}  A_{n'}^{\ell' m_{\ell'}}
\label{eq8}
\eea
\bea
(E_{c} - E) B_E^{\ell'm_{\ell'}} + \sum_{n=1,2} \Omega_n A_{n}^{\ell'm_{\ell'}} (E) = &-& V_{E}^{\ell' m_{\ell'}} - \sum_{\ell m_{\ell}}\left[\int dE' \frac{\mid V^{\ell m_{\ell}}_{E'}\mid^2}{E-E'}\right] B_E^{\ell' m_{\ell'}} \nonumber\\
&-& \sum_{n=1,2} \sum_{\ell m_{\ell}} \left[\int dE' \frac{V^{\ell m_{\ell}}_{E'} \{ \Lambda^{\ell m_{\ell}}_{n}(E')\}^*}{E-E'}\right]A_{n}^{\ell'm_{\ell'}} (E)
\label{eq9}
\eea

Taking $E \rightarrow E + i\eta$ with $\eta=0^+$, we have
\bea
 \sum_{\ell m_{\ell}} \int dE' \frac{V^{\ell m_{\ell}}_{E'} \{\Lambda^{\ell m_{\ell}}_{n}(E')\}^*}{E-E'} = V_{n,eff}(E) - i \pi \sum_{\ell m_{\ell}} \Lambda^{\ell m_{\ell}}_{n}V_{E}^{\ell m_{\ell}}
\label{eq10}
\eea
where
\bea
V_{n, eff}(E) = \sum_{\ell m_{\ell}} {\cal P}  \int dE' \frac{V^{\ell m_{\ell}}_{E'} \Lambda_{\ell m_{\ell}}^{n}(E')}{E-E'}
\eea
is an effective interaction between the closed channel bound state $\mid b_c \rangle$ and the excited bound state $\mid b_e\rangle$ mediated through $s$-wave part of the ground  continuum. Similarly,
\bea
\sum_{\ell m_{\ell}} \int d E' \frac{|\Lambda_{\ell m_{\ell}}^{n}(E')|^2} {E - E'}  = E_{n,pa}^{\rm{shift}} - i \frac{\hbar \Gamma_n}{2}
\label{chio}
\eea
where
\bea
 E_{n,pa}^{\rm{shift}} &=& \sum_{\ell m_{\ell}} {\mathcal P}  \int dE' \frac{|\Lambda^{\ell m_{\ell}}_{n}(E')|^2} {E - E'}\\
 \hbar \Gamma_n &=& 2\pi \sum_{\ell m_{\ell}} |\Lambda^{\ell m_{\ell}}_{n}(E')|^2
\eea
\bea
\sum_{\ell m_{\ell}} \int d E' \frac{\mid V^{\ell m_{\ell}}_{E'}\mid^2}{E-E'} = E_{c}^{{\rm shift}} - i \frac{\hbar \Gamma_f}{2}
\eea
\bea
 \int d E'' \frac{ \mid V_{n, {\rm art}}(E'')\mid^2 }{E - E''} = {\mathcal P} \int d E'' \frac{ \mid V_{n, {\rm art}}(E'')\mid^2 }{E - E''}
 - i \frac{\hbar \gamma_n}{2}
\eea
where
\bea
\gamma_n =  2 \pi  \mid V_{n, {\rm art}}(E)\mid^2
\eea

From Eq.(10) we obtain
\bea
\left(E + \hbar\delta_n - E_{n,pa}^{shift}+ i (\Gamma_n + \gamma_n) /2\right) A_{n}^{\ell'm_{\ell'}} &-& \left(\Omega + V_{n, eff}(E) - i \pi\sum_{\ell m_{\ell}} \Lambda^{\ell m_{\ell}}_{J M_J}V_{E}^{\ell m_{\ell}}\right) B_E^{\ell' m_{\ell'}}\nonumber\\
&=& \Lambda_{n}^{\ell' m_{\ell'}}(E) + {\mathscr K}_{n n'} A_{n'}^{\ell' m_{\ell'}} \hspace{0.8cm} n' \ne n
\label{eq11}
\eea
where
\bea
 {\mathscr K}_{n n'} = \sum_{\ell m_{\ell}} \int dE' \frac{\Lambda^{\ell m_{\ell}}_{n}(E') \{\Lambda^{\ell m_{\ell}}_{n'}(E')\}^*}{E-E'} = E_{n,n'}(E) - i \pi \sum_{\ell m_{\ell}} \Lambda^{\ell m_{\ell}}_{n} (E) \{ \Lambda^{\ell m_{\ell}}_{n'}(E)\}^* - i \pi V_{n, art} (E) V_{n',art}(E)^*
\label{eq10}
\eea
\bea
E_{n n'} = {\cal P} \sum_{\ell m_{\ell}} \int dE'  \frac{\Lambda^{\ell m_{\ell}}_{n}(E') \{\Lambda^{\ell m_{\ell}}_{n'}(E')\}^*}{E-E'}
 +  {\cal P} \int d E'\frac{V_{n,art}(E') V_{n', art}^*(E')}{E-E'}
\eea

Similarly from Eq.(\ref{eq9}) we obtain
\bea
\left(E - E_{c} - E_c^{shift} + i\frac{\Gamma_f}{2}\right) B_E^{\ell'm_{\ell'}} - \sum_{n}\left( \Omega_n +  V_{n,eff}(E) - i \pi \sum_{\ell m_{\ell}}\Lambda^{\ell m_{\ell}}_{n}V_{E}^{\ell m_{\ell}}\right) A_{n}^{\ell'm_{\ell'}} = V_{E}^{\ell' m_{\ell'}}
\label{eq12}
\eea
Define
\bea
\epsilon &=& \frac{E - E_{c} - E_c^{shift}}{\frac{\Gamma_f}{2}}\\
q_n &=& \frac{\Omega_n + V_{n,eff}}{\pi\sum_{\ell m_{\ell}} \Lambda^{\ell m_{\ell}}_{n}V_{E}^{\ell m_{\ell}}}
\eea
From Eq.(\ref{eq12}) we get
\bea
B_E^{\ell'm_{\ell'}} = \frac{V_{E}^{\ell m_{\ell}}}{\frac{\Gamma_f}{2} (\epsilon + i)} + \sum_{n} \frac{\pi\sum_{\ell m_{\ell}} \Lambda^{\ell m_{\ell}}_{n}V_{E}^{\ell m_{\ell}}(q_n-i)}{\frac{\Gamma_f}{2} (\epsilon + i)}A_{n}^{\ell'm_{\ell'}}
\label{eq13}
\eea
and from Eq.(\ref{eq11}) we get
\bea
\left(E + \hbar\delta_n - E_{n,pa}^{shift} + i (\Gamma_n + \gamma_n)/2\right) A_{n}^{\ell'm_{\ell'}} &-& \left((q_n-i)\pi\sum_{\ell m_{\ell}} \Lambda_{n}^{\ell m_{\ell}} V_E^{\ell m_{\ell}}\right) \nonumber\\
&\times& \left[\frac{V_E^{\ell' m_{\ell'}}}{\frac{\Gamma_f}{2}(\epsilon+i)}+\sum_{j=1,2} \frac{\pi\sum_{\ell m_{\ell}} \Lambda^{\ell m_{\ell}}_{j}V_{E}^{\ell m_{\ell}}(q_j-i)}{\frac{\Gamma_f}{2} (\epsilon + i)}A_{j}^{\ell'm_{\ell'}}\right]\nonumber\\
&=&\Lambda_{n}^{\ell' m_{\ell'}}(E) + {\mathscr K}_{n n'} A_{n'}^{\ell' m_{\ell'}}
\eea
Assuming $n=1$ or $n=2$ and $n'=1$ or $n'=2$ but $n' \ne n$, we can express the above equation as a set of two coupled algebraic
equations for $A_{1}^{\ell' m_{\ell'}}$ and $A_{2}^{\ell' m_{\ell'}}$ in the following form
\bea
&& \left(E+\hbar\delta_n - E_{n,pa}^{shift} - E_{qn}^{shift} + i (\ \gamma_n + \Gamma_{qn})/2\right) A_{n}^{\ell'm_{\ell'}} \nonumber \\
&=& \left((q_n-i)\pi\sum_{\ell m_{\ell}} \Lambda_{n}^{\ell m_{\ell}} V_E^{\ell m_{\ell}}\right)
\times \left[\frac{V_E^{\ell' m_{\ell'}}}{\frac{\Gamma_f}{2}(\epsilon+i)}+  \frac{\pi\sum_{\ell m_{\ell}} \Lambda^{\ell m_{\ell}}_{n'}V_{E}^{\ell m_{\ell}}(q_{n'}-i)}{\frac{\Gamma_f}{2} (\epsilon + i)}A_{n'}^{\ell'm_{\ell'}}\right] \nonumber\\
&+&\Lambda_{n}^{\ell' m_{\ell'}}(E) +  {\mathscr K}_{n n'} A_{n'}^{\ell' m_{\ell'}}
\label{twocoup}
\eea
where
\bea
\Gamma_{qn} = \Gamma_n - 2 {\mathrm Im} \left [\frac{\left((q_n-i)\pi\sum_{\ell m_{\ell}} \Lambda_{n}^{\ell m_{\ell}} V_E^{\ell m_{\ell}}\right)^2 }
{\frac{\Gamma_f}{2} (\epsilon + i)}\right ]
\eea
\bea
E_{qn}^{shift} = {\mathrm Re} \left [\frac{\left((q_n-i)\pi\sum_{\ell m_{\ell}} \Lambda_{n}^{\ell m_{\ell}} V_E^{\ell m_{\ell}}\right)^2 }
{\frac{\Gamma_f}{2} (\epsilon + i)}\right ]
\eea
These two coupled algebraic equations can be expressed in a compact form
\bea
A_{n}^{\ell' m_{\ell'}} = \xi_n^{-1} \left [ R_{n}^{\ell' m_{\ell'}} + Q_{n n'} A_{n'}^{\ell' m_{\ell'}}   \right ]
\eea
where
\bea
\xi_n = E+\hbar\delta_n - E_{n,pa}^{shift} - E_{qn}^{shift} + i (\ \gamma_n + \Gamma_{qn})/2,
\eea
\bea
R_{n}^{\ell' m_{\ell'}} = \Lambda_{n}^{\ell' m_{\ell'}}(E) + (q_n-i)\pi\sum_{\ell m_{\ell}} \Lambda_{n}^{\ell m_{\ell}} V_E^{\ell m_{\ell}}
 \frac{V_E^{\ell' m_{\ell'}}}{\frac{\Gamma_f}{2}(\epsilon+i)}
\eea
and
\bea
Q_{n n'} = {\mathscr K}_{n n'} + \left((q_n-i)\pi\sum_{\ell m_{\ell}} \Lambda_{n}^{\ell m_{\ell}} V_E^{\ell m_{\ell}}\right)
\times  \frac{\pi\sum_{\ell m_{\ell}} \Lambda^{\ell m_{\ell}}_{n'}V_{E}^{\ell m_{\ell}}(q_{n'}-i)}{\frac{\Gamma_f}{2} (\epsilon + i)}
\eea
The solution of Eq.(43) yields
\bea
A_{n}^{\ell' m_{\ell'}} = {\mathscr D}_n^{-1} \left [R_{n}^{\ell' m_{\ell'}} + \xi_{n'}^{-1} Q_{n n'} R_{n'}^{\ell' m_{\ell'}}    \right ]
\eea
where
\bea
{\mathscr D}_n = \xi_n - \xi_{n'}^{-1} Q_{n n'} Q_{n' n}
\eea

Now, let us assume that the couplings $V_{E}^{\ell m_{\ell}}$ and $\Lambda_n^{\ell m_{\ell}}$ are real. We further assume that $s$-wave contribution  is most significant such that we can use $V_E^{00} \simeq \sqrt{\Gamma_f/(2 \pi)}  $ and $\Lambda_n^{0 0} \simeq \sqrt{\Gamma_n/(2 \pi)} $. Then d $R_n^{0 0}$ reduces to the standard Fano profile form
\bea
F_n = \frac{\epsilon + q_n}{\epsilon+i}
\eea
Thus we can write
\bea
A_{n}^{0 0} = \frac{F_n \Lambda_n^{00} +  \xi_{n'}^{-1} Q_{n n'} F_{n'} \Lambda_{n'}^{00}}{{\mathscr D}_n }  = \frac{\epsilon + q_n}{\epsilon+i} f_n \Lambda_n^{00}
\eea
where
\bea
f_n = {\mathscr D}_n^{-1} \left [ 1 + \frac{(\epsilon + q_{n'})\Lambda_{n'}}{(\epsilon + q_n)\Lambda_n} \xi_{n'}^{-1} Q_{n n'}   \right ]
\eea

The total inelastic  scattering rate due to spontaneous emission is given by
\bea
\sigma_{inel} = \frac{ \pi }{k^2} \left [ \mid S_{decay} \mid^2 \right ]
\eea
 The average loss rate is given by
\bea
K_{av}^{{\rm res}} = \left \langle v_{rel} \sigma_{inel} \right \rangle
\label{loss}
\eea
where $v_{rel}$ is the relative velocity related to the collision energy by $E = \mu v_{rel}^2/2 = \hbar^2 k^2/2\mu $ and  $\langle \cdots \rangle $ implies
averaging over Maxwell-Boltzmann distribution of the velocity, or equivalently over the collision energy.

\end{document}